\documentclass[twocolumn]{aastex701}

\usepackage{xcolor}

\def\kms{\hbox{km s$^{-1}$}}
\def\VLSR{\hbox{$V_{\rm LSR}$}}
\def\Ekin{\hbox{$E_{\rm kin}$}}

\def\sun{\hbox{$_{\odot}$}}

\def\Msun{\hbox{$M_{\odot}$}}

\begin{document}

\title{Discovery of Multiple Ultra-Broad-Velocity Molecular Features Associated with the W44 Molecular Cloud}

%%%%%%%%%%%%%%%%%%%%%%%%%%%%%%%%%%%%%%%%%%%%%%%%%%%%%%%%%%%%%%%%%%%%%%%%%%%%%%%%

\author[orcid=0009-0003-9269-1899,gname='Momoko',sname='Makita']{Momoko Makita}
\affiliation{School of Fundamental Science and Technology, Graduate School of Science and Technology, Keio University, 3-14-1 Hiyoshi, 4
5 Yokohama, Kanagawa 223-8522, Japan}
\email[]{momoko-m.m@keio.jp} 

\author[orcid=0000-0002-5566-0634,gname='Tomoharu',sname='Oka']{Tomoharu Oka}
\affiliation{School of Fundamental Science and Technology, Graduate School of Science and Technology, Keio University, 3-14-1 Hiyoshi, 4
5 Yokohama, Kanagawa 223-8522, Japan}
\affiliation{Department of Physics, Faculty of Science and Technology, Keio University, 3-14-1 Hiyoshi, Yokohama, Kanagawa 223-8522, Japan}
\email[]{tomo@phys.keio.ac.jp} 

\author[orcid=0000-0002-1663-9103,gname='Shiho',sname='Tsujimoto']{Shiho Tsujimoto}
\affiliation{School of Fundamental Science and Technology, Graduate School of Science and Technology, Keio University, 3-14-1 Hiyoshi, 4
5 Yokohama, Kanagawa 223-8522, Japan}
\email[]{shiho.tsujimoto@keio.jp} 

\author[orcid=0009-0006-9842-4830,gname='Tatsuya',sname='Kotani']{Tatsuya Kotani}
\affiliation{School of Fundamental Science and Technology, Graduate School of Science and Technology, Keio University, 3-14-1 Hiyoshi, 4
5 Yokohama, Kanagawa 223-8522, Japan}
\email[]{sci.tatsu.729@keio.jp} 

%%%%%%%%%%%%%%%%%%%%%%%%%%%%%%%%%%%%%%%%%%%%%%%%%%%%%%%%%%%%%%%%%%%%%%%%%%%%%%%%

\begin{abstract}
We report the discovery of multiple compact molecular features exhibiting extremely broad velocity widths toward the W44 molecular cloud. ALMA CO {\it J\/} = 3--2 data reveal eight ``Petit--Bullets'' surrounding the previously known ``Bullet.'' Each Petit--Bullet shows a distinct V-shaped structure in position--velocity space, reminiscent of the Y-shaped morphology of the Bullet, suggesting a common origin. These features are interpreted as the result of high-velocity plunges of compact gravitational objects into dense molecular gas. The spatial and kinematic properties of the Petit--Bullets suggest that the plunging material was not a single object but rather a small cluster of compact bodies. A virial mass of $1.0\!\times\! 10^{5}\, \Msun$ inferred from their velocity dispersion is comparable to that of typical globular clusters. Momentum analysis further implies that the main Bullet likely formed by an isolated black hole. These findings provide new evidence for dynamical interactions between halo clusters and disk molecular gas.
\end{abstract}



%%%%%%%%%%%%%%%%%%%%%%%%%%%%%%%%%%%%%%%%%%%%%%%%%%%%%%%%%%%%%%%%%%%%%%%%%%%%%%%%

\section{Introduction}
W44 is a ``mixed-morphology'' supernova remnant (SNR), exhibiting a radio continuum shell and centrally peaked X-ray emission \citep[e.g.][]{1978A&A....65L...9G,1985MNRAS.217...99S,1994ApJ...430..757R}.
A radio pulsar is also located within this shell boundary \citep{1991ApJ...372L..99W}.
The distance to W44 is estimated to be 3 kpc based on HI absorption studies \citep{1975A&A....45..239C}.
The remnant is interacting with a neighboring giant molecular cloud \citep[GMC;][]{1998ApJ...505..286S,2004AJ....127.1098S}.
Shocked molecular gas has been traced by 25 OH 1720 MHz maser spots \Citep{1997ApJ...489..143C}, several high-velocity CO wing emissions \citep{2004AJ....127.1098S}, and spatially extended broad CO and HCO$^+$ emissions \Citep{2013ApJ...774...10S}.
Together, these features indicate a thin, expanding shell of shocked gas with an expansion velocity of 13.2 \kms.
Its total kinetic energy, $\Ekin = (1$--$3)\!\times\! 10^{50}$ erg, corresponds to $\sim$10--30\% of the canonical supernova yield.

\begin{figure*}[htbp]
\centering
\includegraphics[width=0.85\linewidth]{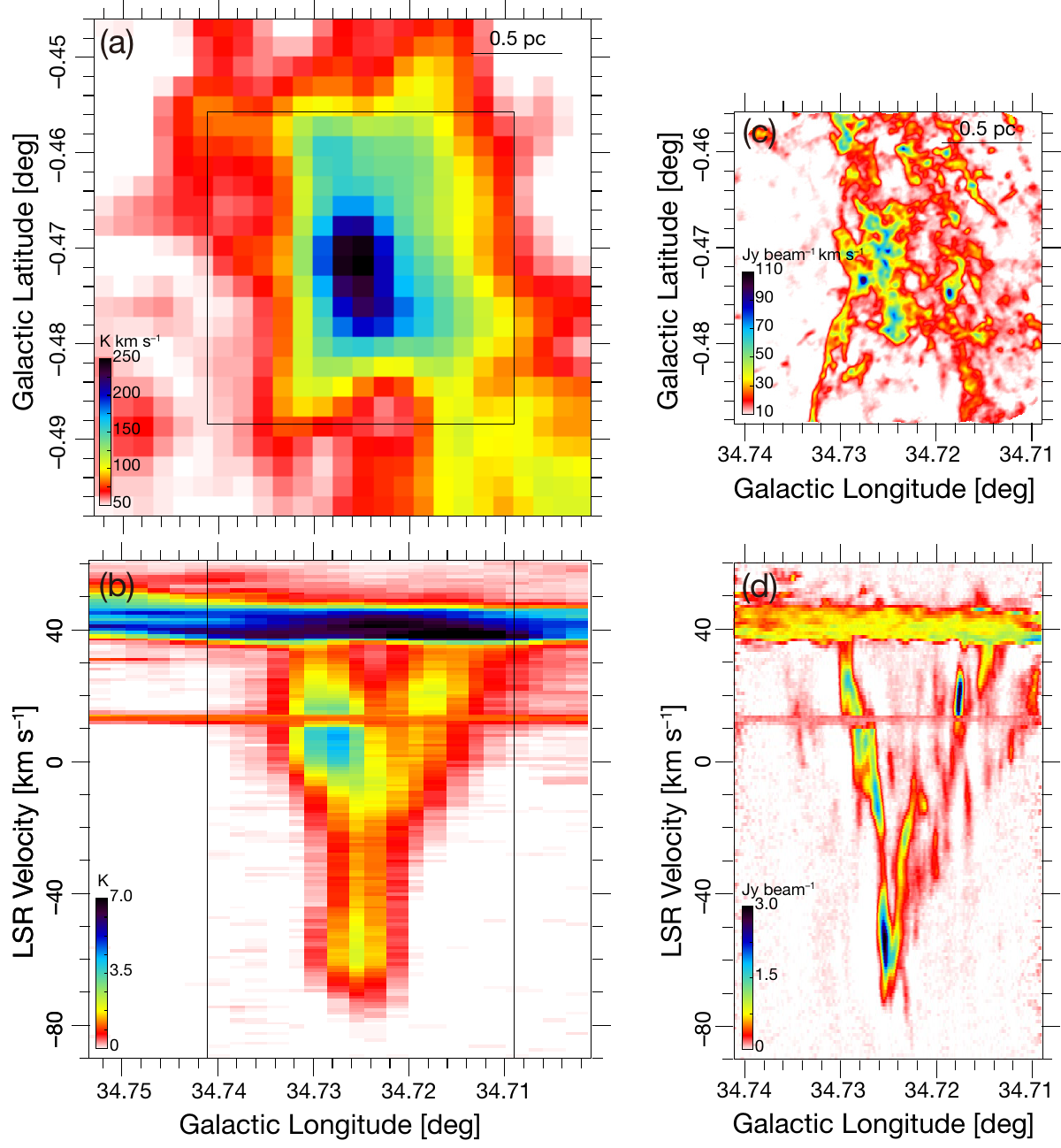}
    % \plotone{fig1}
\caption{(a) Velocity-integrated map and (b) longitude-velocity {\em l--V\/} map of the ASTE CO {\it J\/} = 3--2 data \citep{2017ApJ...834L...3Y}. The integration range is $\VLSR=-90$ to $+60$ \kms, and the intensity is averaged over $b=-0\fdg4716$ to $-0\fdg4722$. The black rectangles indicate the area covered by our ALMA observation.\ (c) Velocity-integrated map and (d) {\em l--V\/} map of the ALMA CO {\it J\/} = 3--2 data. The integration and averaging ranges in velocity and latitude are identical to those in panels (a) and (b), respectively.}\label{fig:int_lv}
\end{figure*}

During a study of the expanding shell kinematics, an ultra-broad-velocity molecular feature was identified at $(l,\,b)=(34\fdg725, -0\fdg472)$, see Figure~\ref{fig:int_lv}b.
This feature is spatially distinct from the W44 pulsar and has a full-width-zero-intensity (FWZI) velocity exceeding 100 \kms, much larger than previously known CO wing emissions ($\lesssim 30\ \kms$).
The emission is detected only at negative velocities relative to the W44 GMC ($\VLSR\sim 40\ \kms$).
Based on its peculiar appearance in the position--velocity map, this feature was named the ``Bullet'' \citep{2013ApJ...774...10S}.
Follow-up observations by \citet{2017ApJ...834L...3Y} have revealed its compact size ($0.5\!\times\! 0.8$ pc) and a characteristic ``Y''-shaped velocity structure.
The Bullet is prominent in CO {\it J\/} = 3--2 and HCO$^+$ {\it J\/} = 1--0 emissions.
Its kinetic energy is $\sim 10^{48}$ erg, about 1.5 orders of magnitude larger than the energy expected from the W44 shock intercepted over its small solid angle.

% %----------Figure Start----------
% %----------Figure End-----------

A plausible scenario of the Bullet is the ``shooting model,'' in which a high-velocity compact object plunges into the dense layer of the W44 GMC. 
Kinetic energy and luminosity considerations suggest that the plunging object has a mass greater than 30 \Msun, 
yet no luminous counterpart is observed, indicating it could be an isolated black hole \citep{2017ApJ...834L...3Y}.
Magnetohydrodynamical simulations reproduce the Bullet's spatial size, kinetic energy, and Y-shaped morphology within this framework \citep{2018ApJ...859...29N}.

Here we report new ALMA observations that reveal additional compact ultra-broad-velocity features surrounding the Bullet, shedding light on the possible clustered nature of the plunging sources.

%%%%%%%%%%%%%%%%%%%%%%%%%%%%%%%%%%%%%%%%%%%%%%%%%%%%%%%%%%%%%%%%%%%%%%%%%%%%%%%%

\section{ALMA Observations}
Observations were performed with ALMA in Cycle 4 (2016.1.01548.S;\@ PI:\@ M.\@ Yamada).
The 12 m array was used on 2017 March 24; the 7 m array on 2017 March 23--24, April 14, and April 16--17; and the total power (TP) array on 2017 March 22--23, April 16--18, and April 22.
The field of view was $0\fdg032\!\times\! 0\fdg033$ centered at $(l,\,b)=(+34\fdg725,\,-0\fdg472)$, covered with 149 pointings of the 12 m array and 53 pointings of the 7 m array.
The 12 m array was in the C40--1 configuration with baseline lengths of 15--155 m.
We targeted CO $J=3$--2 (345.79 GHz) and SiO $J=8$--7 $v=0$ (347.33 GHz) with 2.0 GHz bandwidth and 488 kHz channels.
J1751+0939 and J1924--2914 were used for bandpass and pointing calibrations, J1851+0035 for phase, and J1751+0939 and Titan for flux.

Calibration and reduction were performed using the Common Astronomy Software Applications (CASA).
We used CASA version 4.7.2 to run the calibration script provided by the East Asian ALMA Regional Center.
Continuum emission was subtracted using the task \texttt{uvcontsub}.
Interferometric images were produced with task \texttt{tclean} in natural weighting in CASA version 5.7.0.
The interferometric data were combined with TP data using the Python implementation of the method by \citet{2018AN....339...87F}.
The spatial and velocity grid widths of the resultant cube were $0\farcs13$ and 0.5 \kms, respectively.
The synthesized beam size of the cube was $1\farcs55\!\times\! 1\farcs22$ with a position angle of $-48\fdg0$.

%%%%%%%%%%%%%%%%%%%%%%%%%%%%%%%%%%%%%%%%%%%%%%%%%%%%%%%%%%%%%%%%%%%%%%%%%%%%%%%%

\section{Results}
Throughout this study, the distance to the Bullet is assumed to be 3 kpc, the same as W44 SNR.

\subsection{Overview}
High-resolution CO {\it J\/} = 3--2 data reveal the Bullet's detailed spatial and velocity structure.
Figure~\ref{fig:int_lv}(c) shows the velocity-integrated map over $\VLSR=-90$ to $+60\ \kms$.
The Bullet consists of a main component centered at $(l,\,b)=(34\fdg72, -0\fdg47)$ and filamentary structures extending southeast and northwest.

Figure~\ref{fig:int_lv}(d) presents the $l$--$V$ map averaged over $b=-0\fdg4716$ to $-0\fdg4722$.
A horizontal feature at $\VLSR\sim12\ \kms$ originates from unrelated Galactic spiral arm absorption, while the feature at $\sim40\ \kms$ corresponds to the GMC interacting with W44.
The Y-shaped structure of the Bullet (hereafter referred to as ``Bullet Main''), previously reported by \citet{2013ApJ...774...10S} and \citet{2017ApJ...834L...3Y}, is clearly reproduced over $\VLSR=-80$ to $+40\ \kms$.

\begin{figure*}[htbp]
\centering
\includegraphics[width=0.85\linewidth]{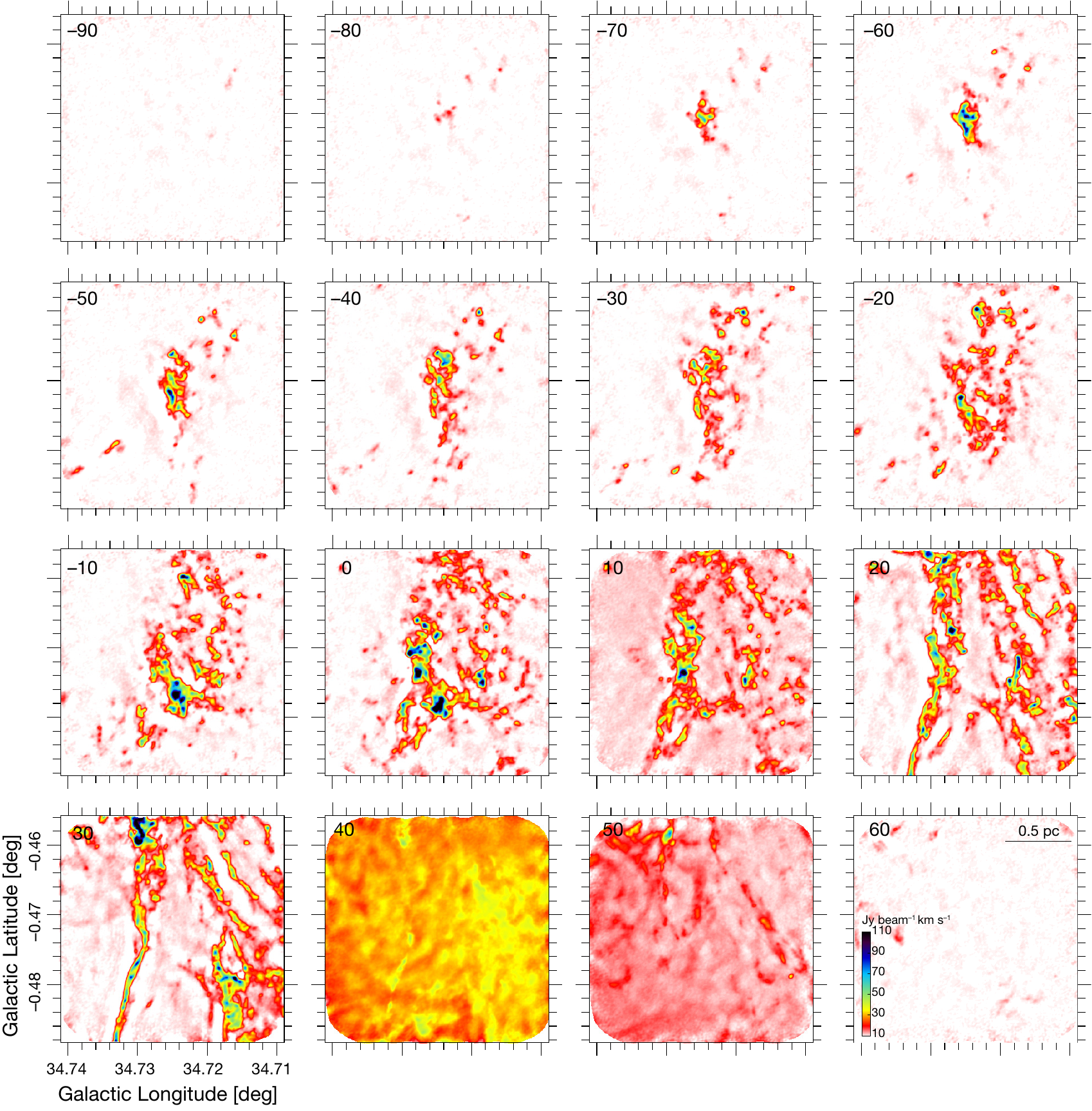}
% \plotone{fig2}
\caption{Velocity channel maps of the Bullet in the CO {\it J\/} = 3--2 line. Each panel shows the integrated intensity over a velocity range of 10 \kms from $\VLSR = -90$ to $+60\ \kms$. The number in the upper-left corner of each panel indicates the central velocity in \kms.}\label{fig:velchan}
\end{figure*}

Figure~\ref{fig:velchan} shows velocity channel maps of the Bullet.
In the $\VLSR\sim 40\ \kms$ panel, faint and diffuse emission from the high-density layer of the W44 GMC appears.
Emission from the Bullet appears only in the panels of $\VLSR\leq 40\ \kms$.
At negative high velocities ($\VLSR=-80$ to $-60\ \kms$), the Bullet Main is concentrated at the center ($\sim0.4$ pc), while it gradually spreads spatially with increasing $\VLSR$.
The filamentary structures toward the southeast and northwest are clearly detected at $\VLSR=+20$ to $+40\ \kms$.

% %----------Figure Start----------
% %----------Figure End-----------

\subsection{Discovery of Petit--Bullets}
In the velocity channel maps, several compact clumps are detected around the Bullet Main. 
Their presence across multiple velocity channels indicates unusually broad line widths. A detailed inspection of the data cube reveals eight new clumps with velocity extents of 74--162 \kms, summarized in Table~\ref{tab:t1}. 
With spatial sizes of 0.1--0.3 pc, they are even more compact than Bullet Main.

\begin{figure*}[htbp]
\centering
\includegraphics[width=0.85\linewidth]{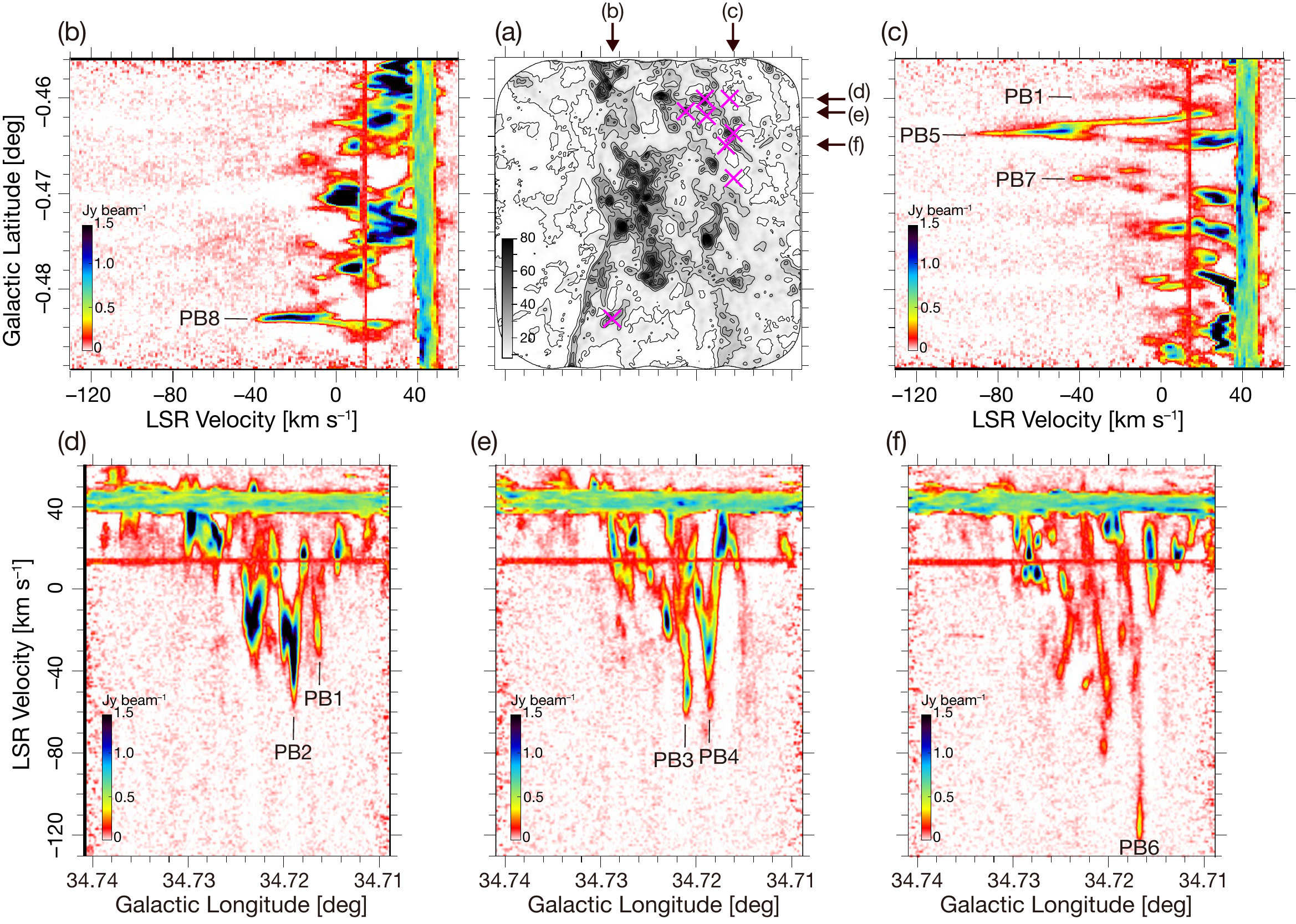}
    % \plotone{fig3}
\caption{(a) Velocity-integrated CO $J$ =3--2 map of the Bullet over $\VLSR=-130$ to $+60\ \kms$. Magenta crosses denote the positions of the Petit--Bullets. Black arrows outside the panel indicate the Galactic coordinates where the position--velocity {\em b--V\/} and {\em l--V\/} maps shown in panels (b)--(f) are extracted.\ (b)--(c) The {\em b--V\/} maps at $l=34\fdg7161$ and $34\fdg7288$ while (d)--(f) The {\em l--V\/} maps at $b=-0\fdg4601$, $-0\fdg4615$ and $-0\fdg4650$, respectively.}\label{fig:pb}
\end{figure*}

Figure~\ref{fig:pb} presents the position--velocity maps of eight clumps.
Each shows a ``V''-shaped feature extending from the W44 GMC at $\VLSR\sim40$ \kms\ toward negative velocities, closely resembling the Bullet Main. 
We therefore designate them as ``Petit--Bullets'' (PBs) and label them PB1--PB8 in order of proximity to the Galactic plane.

% %----------Figure Start----------
% %----------Figure End-----------

%%%%%%%%%%%%%%%%%%%%%%%%%%%%%%%%%%%%%%%%%%%%%%%%%%%%%%%%%%%%%%%%%%%%%%%%%%%%%%%%

\section{Discussions}
\subsection{Physical Parameters of Bullet and Petit--Bullets}
We derived the physical parameters of the Bullet Main and the Petit--Bullets from the CO {\it J\/} = 3--2 data cube.
Angular dispersions along Galactic longitude and latitude, $\sigma_l$ and $\sigma_b$, were measured from the integrated intensity maps and converted into linear scales using $S = D \sqrt{\sigma_l^2 + \sigma_b^2}$ \citep{1987ApJ...319..730S} with $D=3$ kpc.

The velocity width, $\Delta V$, was defined as the difference between the maximum detected velocity (most negative for each clumps) and the systemic velocity of the W44 GMC ($\VLSR = +40\ \kms$). The velocity dispersion, $\sigma_V$, was calculated from the second moment weighted by line intensity. 
The dynamical timescales, estimated as $T_\mathrm{dyn} = S / \sigma_V$, range from a few $\times\! 10^3$ yr for both the Bullet Main and the Petit--Bullets, consistent with a common origin.

The total molecular gas mass, $M_\mathrm{LTE}$, was obtained from the CO {\it J\/} = 3--2 integrated intensity under the assumption of local thermodynamic equilibrium (LTE) and optically thin emission.
An excitation temperature of $T_\mathrm{ex}=40$ K was assumed, derived from the CO {\it J\/} = 1--0, 3--2, and 4--3 data of \citet{2017ApJ...834L...3Y}. A CO-to-H$_2$ abundance ratio of $[{\rm CO}]/[{\rm H_2}]=10^{-4.1}$ \citep{1982ApJ...262..590F} was assumed to convert the CO column density to a molecular hydrogen mass.
The kinetic energy was estimated as $E_\mathrm{kin} = 3/2\, M_\mathrm{LTE} \sigma_V^2$. These parameters are summarized in Table~\ref{tab:t1}.

%-------------------------------Table Start-----------------------------
\begin{deluxetable*}{ccccccccccc}			
\tablenum{1}	
\tablecaption{Properties of the Bullet Main and the Petit--Bullets\label{tab:t1}}	
\tablewidth{0pt}	
\tablehead{
\colhead{Name} 	&	\colhead{$l$} 	&	\colhead{$b$} 	&	\colhead{$\Delta V$}  	&	\colhead{$S$} 	&	\colhead{$\sigma_V$}  	&	\colhead{$T_{\rm dyn}$}  	&	\colhead{$M_{\rm LTE}$}  	&	\colhead{${\rm log}_{10}\,\Ekin$} 	&	\colhead{$M_*$}  	&	\colhead{PA}\\
\colhead{}   	&	\colhead{(degree)} 	&	\colhead{(degree)} 	&	\colhead{(\kms)}  	&	\colhead{(pc)} 	&	\colhead{(\kms)}  	&	\colhead{($10^3$ yr)} 	&	\colhead{(\Msun)} 	&	\colhead{(erg)}  	&	\colhead{(\Msun)} 	&	\colhead{(degree)}	
}
\startdata
Bullet Main 	&	   34.7247 	&	   $-$0.4710   	&	        123.5   &	0.301 	&	26.9 	&	     11.2 	&	7.33 	&	47.2 	&	3.67 	&	\ \,66 	   \\
PB1         	&	   34.7165 	&	   $-$0.4601   	&	   \,\,\,74.0	&	0.090 	&	16.3 	&	\,\,\,5.5 	&	0.22 	&	45.2 	&	0.11 	&	266 	   \\
PB2         	&	   34.7191 	&	   $-$0.4601   	&	   \,\,\,98.0 	&	0.097 	&	23.3 	&	\,\,\,4.1 	&	0.46 	&	45.9 	&	0.23 	&	229 	   \\
PB3         	&	   34.7211 	&	   $-$0.4613   	&	   104.5	    &	0.070 	&	24.1 	&	\,\,\,2.9 	&	0.31 	&	45.7 	&	0.16 	&	149 	   \\
PB4         	&	   34.7189 	&	   $-$0.4619   	&	   \,\,\,93.0  	&	0.090 	&	22.8 	&	\,\,\,4.0 	&	0.36 	&	45.8 	&	0.18 	&	\ \,71 	   \\
PB5         	&	   34.7161 	&	   $-$0.4637   	&	   130.5  	    &	0.108 	&	31.5 	&	\,\,\,3.4 	&	0.41 	&	46.1 	&	0.21 	&	344 	   \\
PB6         	&	   34.7169 	&	   $-$0.4650   	&	   162.0  	    &	0.097 	&	35.6 	&	\,\,\,2.7 	&	0.39 	&	46.2 	&	0.20 	&	110 	   \\
PB7         	&	   34.7161 	&	   $-$0.4684   	&	   \,\,\,84.5 	&	0.106 	&	19.2 	&	\,\,\,5.5 	&	0.26 	&	45.5 	&	0.13 	&	\,\ \ 7    \\
PB8         	&	   34.7288 	&	   $-$0.4831   	&	   \,\,\,81.0	&	0.132 	&	16.8 	&	\,\,\,7.8 	&	0.75 	&	45.8 	&	0.37 	&	239 	   \\
\enddata
\end{deluxetable*}
%-------------------------------Table End------------------------------

\subsection{Determination of Velocity Vector}\label{subsec:vector}
\begin{figure*}[htbp]
\centering
\includegraphics[width=0.85\linewidth]{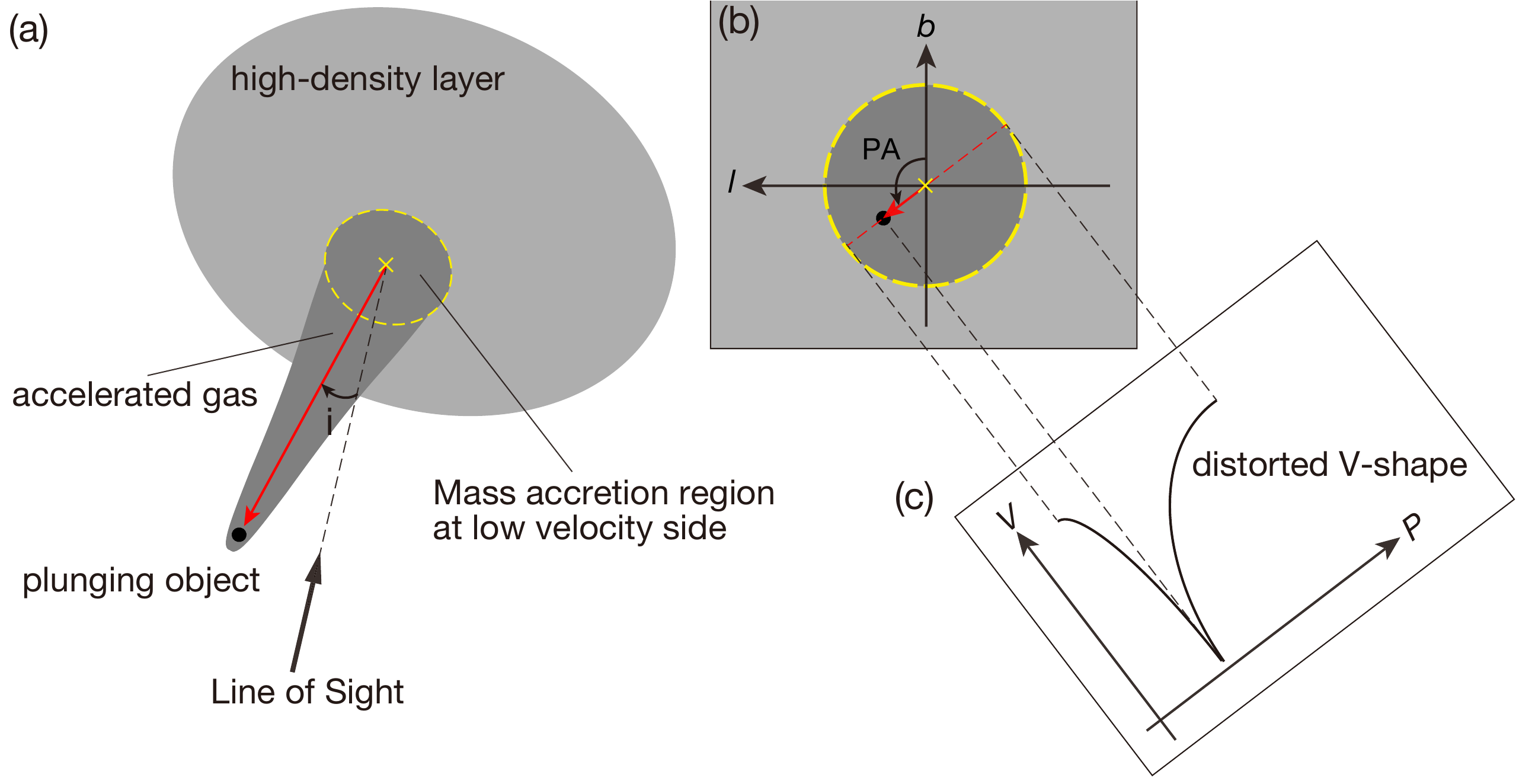}
% \plotone{fig4}
\caption{(a) Schematic of the shooting model in which the plunging object moves obliquely with respect to the line of sight. Red arrow indicates the plunge direction. The yellow dashed circle marks the low-velocity accretion region centered at the yellow cross, and the black dot represents the high-velocity peak.\ (b) $l$--$b$ map viewed along the line of sight, corresponding to the observed morphology for an oblique plunge.\ (c) Position--velocity diagram cut along the plunge direction (red arrows), showing a distorted Y-shaped feature.}\label{fig:vel_vector}
\end{figure*}

The similarity in the position--velocity structures of the Bullet Main and the Petit--Bullets may indicate a common origin.
We thus use the framework of the shooting model proposed by \citet{2017ApJ...834L...3Y} to interpret their formation. In the original model, this plunge is assumed to be perpendicular.
In practice, however, the observed Y- and V-shaped features are not perfectly symmetric but rather distorted. 
As shown in Figure~\ref{fig:vel_vector}, this asymmetry can be reasonably interpreted as a projection effect, indicating that the actual plunge was oblique relative to the line of sight ($i \neq 0$ in panel~a).

To provide a tentative quantification, we identified the centroid of the low-velocity, mass-accretion region, which likely corresponds to the initial impact site.
The high-velocity peak represents the current position of the compact object.
The spatial offset between these points was used to estimate the projected plunge direction, expressed as the position angle (PA; see Figure~\ref{fig:vel_vector}b), and the associated velocity component.

To convert this offset into a velocity, the elapsed time $\Delta t$ since the impact must be estimated. Following \citet{2018ApJ...859...29N}, we assume that the root width $d$ of the Y- and V-shaped features scales with the Alfv\'en velocity $v_A = B_{x0}/\sqrt{4\pi \rho_0}$ and $\Delta t$. Adopting representative values of $B_{x0} = 500\ \mu$G \citep{2005ApJ...627..803H} and $n=10^4\ \mathrm{cm}^{-3}$ \citep{2017ApJ...834L...3Y}, we obtain $v_A \approx 11\ \kms$. This yields an order-of-magnitude estimate of $\Delta t$ for each Bullet, allowing us to infer the projected plunge velocities. We note that local inhomogeneities in the gas density and magnetic field may introduce additional scatter in the observed sizes and morphologies.

The resulting position angles are summarized in Table~\ref{tab:t1} together with other physical parameters. While the line-of-sight velocities show a coherent negative trend, the position angles on the plane of the sky are largely random.
The average velocity components in the sky plane were found to be $V_l \sim 0.8\ \kms$ and $V_b \sim -0.6\ \kms$, with dispersions of $\sigma_{V_l} \sim 10.4\ \kms$ and $\sigma_{V_b} \sim 12.4\ \kms$.
We note, however, that these estimates are subject to significant uncertainty, as the centroid determinations are affected by contamination from surrounding structures.

% %----------Figure Start----------
% %----------Figure End-----------

\subsection{Nature of the Plunging Objects}
We discuss the nature of the plunging objects responsible for the formation of  the Bullet Main and the Petit--Bullets.
These structures are interpreted to have been formed by plunging compact objects. 
Because the features are spatially concentrated within $\sim1$~pc (in projection), they are likely composed of a small group of compact objects rather than a single object.
The inferred plunging velocities, which are opposite to the Galactic rotation ($\sim40~\mathrm{km~s^{-1}}$), reach 74--162 \kms.  
Such high and counter-rotating velocities suggest that the plunging objects may have originated from the Galactic halo rather than from the disk.

No stellar or bright counterparts are detected at other wavelengths \citep[][]{2006AJ....131.1163S,2017ApJ...834L...3Y,2016A&A...595A...1G,2023A&A...674A...1G}, 
implying that the plunging objects are unlikely to be massive stars. 
For each structure, we estimated the line-of-sight momentum of the accelerated gas from its mass and velocity distribution, and derived a lower limit on the mass of plunging objects ($M_*$) based on momentum conservation (Table~\ref{tab:t1}).  
Here, we assume that the velocity of each plunging object ($V_*$) corresponds to the velocity width ($\Delta V$) of the Y- or V-shaped features and that the object's entire momentum is transferred to the gas.  
Therefore, the derived mass represents a conservative lower limit.
The lower limit for the Bullet Main exceeds the Tolman-Oppenheimer-Volkoff (TOV) limit \Citep[1.5--3 \Msun;][]{2001ApJ...550..426L}, suggesting the presence of an isolated black hole.
In contrast, the lower limits for the Petit--Bullets can be explained by stellar-mass objects, whose potential counterparts may have remained obscured at optical and near-infrared wavelengths in previous observations due to strong interstellar extinction.

Furthermore, by combining the sky-plane velocity dispersion derived in
Section~\ref{subsec:vector} with the line-of-sight velocity dispersion
of the plunging objects ($\sigma_{V_*}=27\ \kms$), we obtain a
three-dimensional velocity dispersion of
$\sigma_\mathrm{3D}=31.5\ \kms$.
Assuming a spherical geometry and adopting the Bullet's spatial extent,
we estimate a characteristic virial mass scale using the simple virial
relation $M_\mathrm{VT} \sim R\,\sigma_\mathrm{3D}^2 / G$, where
$R\ (\sim0.45\ \mathrm{pc})$ represents the projected size of the
plunging-object distribution. The resulting mass scale,
$M_\mathrm{VT}\sim 1.0\times10^5\ M_\odot$, should be regarded as an
order-of-magnitude estimate of the dynamical scale rather than a
precise mass determination.
This value is comparable to that of typical globular clusters, which
provides one possible reference for contextualizing the inferred
dynamical scale of the system.

%%%%%%%%%%%%%%%%%%%%%%%%%%%%%%%%%%%%%%%%%%%%%%%%%%%%%%%%%%%%%%%%%%%%%%%%%%%%%%%%

\section{Summary}
We report the discovery of eight compact, high-velocity clumps, ``Petit--Bullets,'' surrounding the previously known ``Bullet'' in the W44 supernova remnant, revealed by high-resolution ALMA CO {\it J\/} = 3--2 observations. 
These clumps are spatially compact (0.1--0.3 pc) and exhibit broad V-shaped velocity widths. Their kinematic and morphological characteristics closely resemble those of Bullet Main, suggesting a common origin. 
We interpret this within an extended ``shooting model'' \citep{2017ApJ...834L...3Y}, in which a group of point-like gravitational sources plunges into the dense molecular layer. 
The plunging object associated with the Bullet Main is inferred to be an isolated black hole, whereas the Petit--Bullets can be explained by stellar-mass objects. 
Future ALMA observations using higher-density tracers (e.g., HCO$^+$) will refine the velocity vectors and reveal their internal gas structures that remain unresolved in the current CO {\it J\/} = 3--2 data.
Deep near-infrared imaging will test this scenario by searching for possible stellar counterparts to the Petit--Bullets. These efforts will constrain the physical nature and origin of the Bullet, potentially unveiling a previously unrecognized class of compact, high-velocity stellar objects (remnants).

%%%%%%%%%%%%%%%%%%%%%%%%%%%%%%%%%%%%%%%%%%%%%%%%%%%%%%%%%%%%%%%%%%%%%%%%%%%%%%%%

\begin{acknowledgments}
This paper makes use of the following ALMA data: ADS/JAO.ALMA\#2016.1.01548.S. ALMA is a partnership of ESO (representing its member states), NSF (USA) and NINS (Japan), together with NRC (Canada), MOST and ASIAA (Taiwan), and KASI (Republic Korea), in cooperation with the Republic of Chile. The Joint ALMA Observatory is operated by ESO, AUI/NRAO and NAOJ.
T.O. acknowledges the support from JSPS Grant-in-Aid for Scientific
Research (A) No. 20H00178.
\end{acknowledgments}

\facility{ALMA.}

\software{CASA \citep{2022PASP..134k4501C}, Matplotlib \citep{2007CSE.....9...90H}, NumPy (\citealp{2011CSE....13b..22V, 2020Natur.585..357H}), and SciPy \citep{2020NatMe..17..261V}.}

%%%%%%%%%%%%%%%%%%%%%%%%%%%%%%%%%%%%%%%%%%%%%%%%%%%%%%%%%%%%%%%%%%%%%%%%%%%%%%%%

% \appendix

% \section{Appendix information}

%% For this sample we use BibTeX plus aasjournalv7.bst to generate the

\bibliography{reference}
\bibliographystyle{aasjournalv7}

\end{document}